# Optimizing the walk coin in the quantum random walk search algorithm through machine learning

**Hristo Tonchev** [1]*

[1] Institute of Solid State Physics, Bulgarian Academy of Sciences,72 Tzarigradsko Chaussée, 1784 Sofia, Bulgaria

\* Correspondence: htonchev@issp.bas.bg;

**Petar Danev** [2]

[2] Institute for Nuclear Research and Nuclear Energy, Bulgarian Academy of Sciences,72 Tzarigradsko Chaussée, 1784 Sofia, Bulgaria

**Abstract:** This paper examines the stability of the quantum random walk search algorithm, when the walk coin is constructed by generalized Householder reflection and additional phase shift, against inaccuracies in the phases used to construct the coin. The optimization of the algorithm is done by numerical methods - Monte Carlo, neural networks, and supervised machine learning. The results of numerical simulations show that, with such a construction of the Householder reflection, the algorithm is more stable to inaccuracies in the specific values of these phases, as long as it is possible to control the phase difference between the phase shift and the phase involved in the Householder reflection. This paper explicitly shows as an example, how achieving a properly designed phase difference would make quantum random walk search on a hypercube more stable for coin register consisting of one, two, and three qubits.

1. Introduction

Quantum mechanics allows the quantum analogue of the bit - qubit to be both in pure state and in superposition of basis states. Quantum superposition, entanglement, the evolution of this superposition, and the collapse of the wave function in measurement allow the quantum computer to efficiently solve problems that the classical computer could not [1]. These phenomena give quantum computers a number of advantages over classical ones - the ability to accelerate many algorithms [1], potential to create secure cryptographic protocols [2], effective simulations of quantum systems [3] and other. Mathematically, the problems that are solved efficiently by the quantum computer are of the computational class BQP [4], while the classic ones are of the computability class BPP. The class in which the problems solved effectively on the classical computer is a subclass of the problems solved effectively on the quantum one.

Classic computers have been improved for many years [5]. In order for quantum computer to prevail, it is necessary to catch up with this advantage and show that it can solve certain problems faster than the classic ones - to achieve quantum superiority. So far, quantum superiority has been demonstrated in solving several problems [6], [7].

There are various physical systems that could be used to experimentally implement a quantum computer. Examples of such systems are based on nuclear magnetic resonance of the macromolecules [8], nitrogen defects in a diamond crystal lattice [9], quantum dots [10], ions in the ion trap [11], atoms in an optical lattice [12], superconducting vortices [13] and Josephson junctions [14]. In order for a quantum system to be used to create a quantum computer, it must meet DiVincenzo's criteria [15]. Additional challenges come from the requirement that the quantum computer should be more efficient than the classical ones. Each physical realization of quantum computers meets these criteria differently and has its own advantages and disadvantages [16]. In those realizations, the operations on the qubits are made through different interactions. This makes it possible to easily implement some gates on specific physical systems. An example of this is the Householder reflection [17], which can be easily implemented in ion traps. It can be used to build quantum gates more efficiently [18], [19]. Constructing these gates requires optimization of various parameters, which include those of the specific laser systems used. Many groups are currently working on improving this and other implementations.

The advantages of quantum computers over classical ones are the reason for the increased interest in both the scientific community and many private companies. Quantum computers developed by IBM and Google are based on Josephson junctions [14]. IBM announced in early 2020 that they have 28-qubit quantum computers and a 53-qubit system under development [20]. Google announced in the late 2019 that they have a programmable quantum computer with 54 qubits [6]. Another leading company in the development of quantum computers is Intel, which have implemented computers with 49 superconducting qubits [21].

Different quantum algorithms have different accelerations compared to their classical analogues. Acceleration in quantum Shor's factorization algorithm [1] and the quantum algorithm for calculating a discrete logarithm [1] is exponential. These two quantum algorithms can break classical cryptographic protocols RSA [22] and the cryptographic protocol of elliptic curves [23]. Quantum search algorithms in an unsorted database [1], [24], [25] and the algorithm for calculating the Boolean formulas [26] have a quadratic acceleration compared to their classical analogues. The acceleration of the quantum algorithm for searching in a sorted database has a constant multiplier [27].

Quantum algorithms are conditionally divided into several categories based on the operation that underlies the algorithm and can be performed efficiently on the quantum computer. Some algorithms fall into more than one category. Only a few such categories will be mentioned here. Examples of algorithms based on amplitude amplification [28], [29] include Grover's algorithm [1], [19] and quantum counting algorithm [1], [30]. Quantum Fourier transform [31] is the basis of Shor's algorithm [1], phase estimation algorithm [30] and quantum counting algorithm [1], [30], which also falls into this category. Algorithms based on quantum random walk [32] are the third major category, and this category itself is divided into two sub-categories - continuous time quantum random walk [33] and discrete time quantum random walk [34].

There are different quantum algorithms for searching in an unsorted database, examples are the Grover's algorithm [1] and discrete time quantum random walk search (QRWS) algorithm [24]. These algorithms are quadratically faster than classical algorithms, but have various advantages and disadvantages. This makes them suitable for different tasks. Grover's algorithm contains two times less oracle calls compared to quantum random walk-based ones, but can only be used to search in a linear database. The random walk search algorithm can be used for a database with any topology.

Discrete time quantum random walk was proposed by Aharonov *et al.* in 1993 [34]. Initially, they used it to study simple structures, such as a line [35] and a circle [35]. The quantum walk algorithm is quadratically faster than the classical ones. The faster traversal of graphs is the reason why it is used to transverse other structures, such as square [36], hexagonal grid [37], torus [36], hypercube [38], and so on. The quantum walk has been used as a basis for making various quantum algorithms. Examples of such algorithms are the quantum algorithm for finding distinct elements [39], the quantum algorithm for finding triangles in a graph [40], quantum algorithm for calculating Boolean formulas [41], and the quantum random walk search algorithm [24].

The quantum random walk search algorithm was first introduced by Shenvi *et al.* in 2008. [24]. There are different variants of the algorithm designed to search on different topologies. An example of topologies of great interest are tree graph [42], square grid [36], hypercube [43], regular d-dimensional grid [44] and fractal structures [45].

Some modifications of the random walk search on a hypercube increase the probability of finding the wanted element. Examples of this is the modifications of the shift operator [43], search on the nearest neighbors [43], and coin modification [46]. For a hypercube, the case in which the algorithm has more than one solution is also studied and it is shown how to calculate the number of iterations in this case [47].

In the paper we conduct detailed numerical study and optimization of the Quantum random walk search algorithm with Monte Carlo and Machine learning methods.

Monte Carlo methods are stochastic methods [48], [49]. Their application begins with presenting the problem as a mathematical function depending on one or more parameters. The function is solved numerically by taking random values of the parameters and the results of the computation are recorded. The methods are suitable for probabilistic and deterministic problems, and the function itself can be solved repeatedly to accumulate sufficient results. They are widely used in solid state physics [50], high energy physics [51], molecular biology [52], computational ecology [53] and many others.

Machine learning is a class of numerical algorithms that build a model based on sample data, known as training data, in order to identify patterns, make predictions, or make decisions on its own [54]. It is a sub-field of artificial intelligence. Machine learning could be divided in three main types: unsupervised [55], supervised [56], and reinforced [57]. The unsupervised learning algorithms' input is unlabeled training data. They classify, categorize, and create clusters from the input data. The supervised learning use labeled training examples to learn patterns and to create a mathematical model that maps the input data to the labeled output data. It can be used in classification or, as in this paper – in regression problems. The reinforcement machine learning methods use agents that learn by interacting with other agents or with the environment, receive rewards when making correct decisions, and their behavior is based on maximizing the reward. It is not expected for the agents to have any prior knowledge for the problem they are expected to solve.

Machine learning finds application in quantum mechanics in the design of algorithms and experiments that are difficult to calculate with analytical methods [58] and in the processing of huge databases with experimental results [59]. Examples of quantum algorithms designed in this way are the quantum algorithm for solving inhomogeneous linear partial differential equations [60], Deutsch–Jozsa algorithm with oracle constructed by supervise learning [61], the quantum error correction algorithm [62], and so on.

This article is organized as follows: Chapter 2 briefly discusses the quantum random walk search algorithm. In Chapter 3 the walk coin's parameters of QRWS algorithm that will be optimized by Monte Carlo and Machine Learning methods are elaborated. First in 3.1, it is shown how to find the optimal parameters on an arbitrary graph with equal probability for choosing any direction from all available, being at the current vertex. In 3.2. it is shown how to obtain a coin corresponding to these parameters using a generalized Householder reflection and a phase multiplier. The last part of the third chapter underlines the importance of these parameters for quantum random walk search on a hypercube. Chapter 4 discusses the numerical methods used in our work - namely Monte Carlo 4.1 and Machine Learning 4.2. The next part of the chapter 4.3 explains how an optimal for the set tasks neural network has been constructed. In Chapter 5 are presented the main results of our work. The results of the Monte Carlo simulations are shown first in 5.1. They are used to find walk coin parameters that give high probability to find the searched element and are resistant to inaccuracies in these parameters. A more optimized coin solutions, through a combination of Monte Carlo and machine learning, are given in 5.2. The results of numerical simulations are shown for a coin consisting of one 5.2.1., two 5.2.2. and three 5.2.3. qubits. Machine learning was used to hypothesize an optimization for a coin composed of a larger number of qubits. Brief remarks on experimental realizations of the Householder reflections and phase gate in ion traps are given in 6. The conclusion of this work is in Chapter 7.

## 2. Quantum random walk search algorithm

The quantum random walk search (QRWS) algorithm can be used to search on structures with any topology. The steps of the algorithm will be given, alongside with example of quantum random walk search on a hypercube (QRWSoH). The algorithm's quantum circuit is shown on FIG. 1.

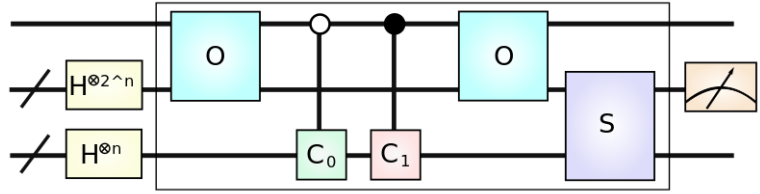

FIG. 1: *Quantum circuit of the QRWS algorithm. With $H$ is denoted the Hadamard gate, with $O$ the oracle, with $S$ the shift operator. The two coins, the traversal and the marking ones, are denoted by $C_0$ and $C_1$.*

Each $D_L$ - dimensional cube consists of $E_{D_S,D_L}$ number of cubes of dimension $D_S$, where $D_L \geq D_S$. Important special cases are the number of vertices (nodes) of $D_L$- dimensional cube $E_{0,D_L} = 2^{D_L}$ and the number of edges at each vertex of the hypercube (the number of states of the coin register) $E_{1,D_L} = D_L$. Quantum registers are built of qubits, therefore $D_L$ must be a power of two. If the number of qubits of the coin is n, then $D_L = m = 2^n$. For example, for a coin of 3 qubits the algorithm has edge (coin) register consisting of $E_{1,D_L} = 8$ states and the register of vertices will be of eight qubits with $E_{0,D_L} = 256$ states.

Hamming weight is the number of characters by which a string differs from a string with the same number of elements but filled only with zeros. Each vertex of the hypercube can be numbered by using digits in a binary number system FIG. 2, so that the movement in each direction changes by one the Hamming weight. Probability to go in each direction is 1/m.

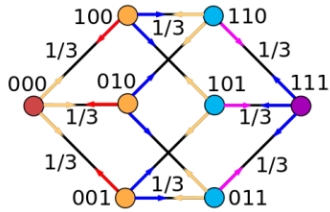

FIG. 2: *Numbering of vertices of hypercube in quantum random walk on a hypercube. Probability to go from each edge to any of its neighbors is 1/3.*

The quantum random walk search QRWS algorithm begins with preparation of the initial states of the register of graph nodes and coin register in an equal-weight superposition of all states. This can be done by applying Hadamard gate $H$ to each of the qubits in both registers. On FIG. 1 applying $H$ to register of the coin consisting of $n$ qubits is denoted by $H^{\otimes n}$, and applying the same gate to the register of nodes consisting of $2^n$ qubits is denoted by $H^{\otimes 2^n}$. The number of states in both registers are respectively $E_{1,D_L} = m = 2^n$ and $E_{0,D_L} = 2^{2^n}$.

The iteration $W$ of the algorithm must be applied $k_{itr}$ times to the initial state prepared in this way.

$$k_{itr} = \left\lceil \frac{\pi}{2}\sqrt{2^{m-1}} \right\rceil \tag{1}$$

The brackets $\lceil\ \rceil$ denote rounding up the number of steps.

Each iteration $W$ of the algorithm consists of the following steps:

1) An oracle $O$ is applied.

    The oracle is an operator that acts on the vector of nodes and each solution corresponds to a specific vertex. The operator is built around function $f(\vec{x})$ that cannot find the solution, but can recognize it. In many problems the solution in one direction is simple and in the opposite direction is with higher complexity. A set of solutions will be denoted by $\Omega$. Action of the function on each state of node register $\vec{x}$ can be written as:

$$f(\vec{x}) = \begin{cases} 1 & \vec{x} \in \Omega \\ 0 & \vec{x} \notin \Omega \end{cases} \tag{2}$$

    The oracle marks the searched state and thus determines which of the two coins to be applied to a particular node state.

2) The respective coin is applied.

    a) The traversing (walking) coin $C_0$ is used on the unmarked nodes. It determines the probability for choosing any direction from all available, being at the current vertex. Since the probability of going in each direction should be the same, the coin must have the following form:

$$C_0 = \begin{pmatrix} a & b & \cdots & b \\ b & a & \cdots & b \\ \vdots & \vdots & \ddots & \vdots \\ b & b & \cdots & a \end{pmatrix} \tag{3}$$

    An optimal coin for traversing a hypercube is the Grover's coin. It is a real matrix with a dimension $m$, which has the form of $C_0$, and its matrix elements are determined by the equations:

$$a = -1 + 2m^{-1} \qquad\qquad b = 2m^{-1} \tag{4}$$

    b) The coin $C_1$ is applied to the marked states. The negative identity operator is used for the marking coin.

$$C_1 = -I = \begin{pmatrix} -1 & 0 & \cdots & 0 \\ 0 & -1 & \cdots & 0 \\ \vdots & \vdots & \ddots & \vdots \\ 0 & 0 & \cdots & -1 \end{pmatrix} \quad (5)$$

Both coins have same dimension $Dim(C_1) = Dim(C_0) = E_{1,D_L}$.

3) The oracle $O$ is applied a second time.
4) The shift operator $S$ is applied.

Different topologies are determined by the shift operator, which defines the edges connecting the nodes on the graph. The shift operator changes the state of the node register depending on the state of the coin register. The shift operator for quantum random walk on a hypercube is:

$$S = \sum_{i=0}^{m-1} \sum_x |i, x\rangle \langle i, x^i|, \quad (6)$$

where $x^i$ is the state $x$ with its $i$-th bit flipped.

After making the required number of iterations $k_{itr}$, the state of the register of the graph's vertices is measured. With some probability the measured state is the searched state. It is checked whether the measured state is the required one. If not, the algorithm is repeated.

A key feature of the QRWS algorithm is that the state of the vertex register must be measured at the exact moment. During execution of the algorithm, both nodes and edge registers are in superposition of states. At the moment of measurement of the register's state, superposition collapses into a pure state. If at each step of the algorithm the state of the vertices and/or the state of the coin is measured, then the quantum walk collapses to classical one.

The probability of finding the solution after measurement is a periodic function of the number of applications of the operator $W$. It has period of $\pi\sqrt{2^{m-1}}$. The period, unlike the number of iterations, is not an integer. This leads to the need to round the number of iterations, so probability of finding the searched element differs slightly at each maximum given by the expression $\left[(2k+1)\frac{\pi}{2}\sqrt{2^{m-1}}\right]$, where $k$ is an integer.

On FIG. 3 are shown the results of numerical simulations of the QRWS algorithm for a hypercube, and the Grover's coin used for traversing coin. FIG. 3 a) shows the probability to obtain each of the states of node register, after the measurement. Node register is measured after $k_{itr}$ iterations. The searched element is the 2nd. The coin register has 4 states and the node register has 16 states. An example of the change in the probability of finding the searched element from the number of iterations of algorithm for hypercube topology is shown on FIG. 3 b). The number of states of the coin (Hilbert space of the coin) is 8 (3 qubits), and number of states of the nodes (Hilbert space of nodes) is 256 (8 qubits). Probability to find the element marked by the coin does not depend on which element the coin marks.

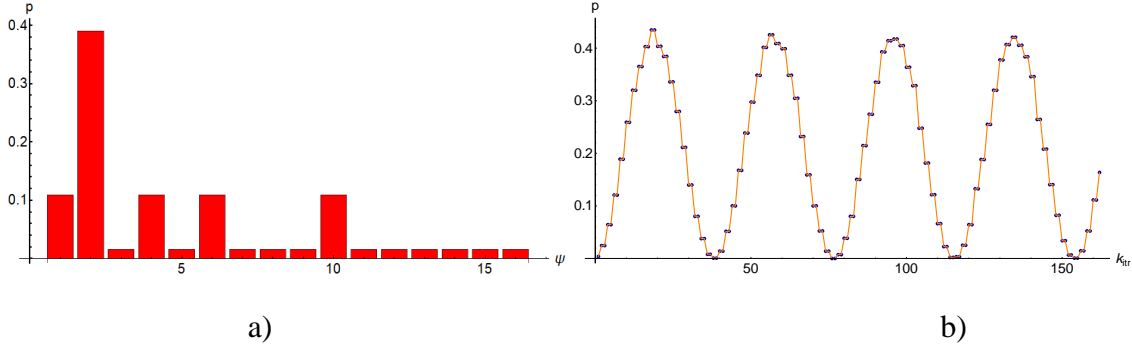

FIG. 3: *Numerical simulations of QRWSoH. Searched element is the second one. Figure a) shows the probability of finding each element of the node register after the required number of iterations. The states of the coin are 4 and states of the nodes - 16. On figure b) is shown the probability of finding searched element as a function of the number of iterations. The number of states of the vertices is 256, and accordingly the states of the coin are 8.*

The probability of finding the searched element is approximately $1/2 - O(1/2^m)$, where $m$ is the number of states of the vertex register.

### 3. Determining the optimization parameters

In this section the optimization of the traversing coin used in the quantum random walk search algorithm will be discussed. This is important because it would improve the probability to find solution and the speed of the algorithm. Appropriate optimization parameters will be found first and it will be shown how they can be obtained by Householder reflection. At the end of this section, these parameters will be considered in the context of the hypercube topology.

The optimization will be conducted by Monte Carlo (MC) and machine learning (ML) methods. First, MC simulations will select random samples of these parameters within certain limits and generate datapoints that will be used by the machine learning algorithm. In order to use ML, it is extremely important to determine the input parameters for the neural network (NN), which coincide with the optimization parameters of the traversing coin. The smaller the number of these parameters, the more universal will be the result obtained for the corresponding algorithm based on quantum random walk. A small number of parameters would also reduce the required number of datapoints needed to obtain results and would help to quickly find the required neural network configuration. A lower number of optimization parameters leads to low number of neurons needed in each layer as well as to fewer layers of the NN, which in turn would increase the speed at which the neural network will be trained and will reduce the memory required for training the model.

### 3.1. Traversing operator and search on any regular graph

When the probability for choosing any direction from all available is equal, then each traversing coin has a matrix representation with one value for each element on the main diagonal and another value for each element outside it.

$$C_0 = \begin{pmatrix} a & b & \cdots & b \\ b & a & \cdots & b \\ \vdots & \vdots & \ddots & \vdots \\ b & b & \cdots & a \end{pmatrix} \tag{7}$$

The matrix $C_0$ has dimension $m$. Matrix elements $a$ and $b$ can be complex numbers. It follows from the requirement for the reversibility of quantum calculations that the matrix is unitary - $C_0 C_0^\dagger = I$.

The optimal traversal of any regular graph (and in particular of the hypercube) is done with a coin that can be found using the unitarity of the matrix [63]:

$$|a|^2 + (m-1)|b|^2 = 1 \qquad\qquad 2|a||b|\cos\Delta + (m-2)|b|^2 = 0 \tag{8}$$

If the matrix elements are expressed with amplitude and phase, scilicet $a = |a|\exp(-i\varphi_a)$ and $b = |b|\exp(-i\varphi_b)$, then the phase-difference $\Delta = \varphi_a - \varphi_b$ is

$$\cos\Delta = \frac{1 - (m-2)|b|^2}{2|a||b|} \tag{9}$$

Ignoring the trivial solution $|a| = 1$ and $|b| = 0$, for the moduli of $a$ and $b$, where the coins' register is bigger than two ($m > 2$) the following expressions are obtained:

$$|a| = \frac{-2 + m}{\sqrt{2 + (-2+m)m + 2(-1+m)\cos(2(\varphi_a - \varphi_b))}} \tag{10}$$

$$|b| = -\frac{2\cos(\varphi_a - \varphi_b)}{\sqrt{(-2+m)^2 + 4(-1+m)\cos(\varphi_a - \varphi_b)^2}} \tag{11}$$

When only random walk is considered, it is only the phase difference $\varphi_a - \varphi_b$ that matters, and not the specific values of the two angles. The phase difference between the angles leads to a phase multiplier in front of the whole coin which is unobservable and can be ignored. However, for the QRWS, the phase multiplier in front of the whole coin cannot be ignored.

## 3.2. Householder reflection and search on a regular graph

The generalized Householder reflection $M$, given by

$$M(\phi, \chi) = I - (1 - e^{i\phi})|\chi\rangle\langle\chi| \qquad (12)$$

is widely used in quantum information. $\phi$ is a phase, and for the state vector $|\chi\rangle$ should be used equal weight superposition of basis vectors $|i\rangle$ of the coin register. Remarks on easy implementation with ion traps of the coin by Householder reflection will be given at the end of this work.

Each operator can be decomposed efficiently by using Householder reflections [17]. The matrix elements of the traversing coin $C_0$ for a regular graph can also be obtained by only one generalized Householder reflection and additional phase multiplier $e^{i\varsigma}$. The traversing coin can be written as:

$$C_0'(\phi, \chi, \varsigma) = e^{i\varsigma} M(\phi, \chi) = e^{i\varsigma}(I - (1 - e^{i\phi})|\chi\rangle\langle\chi|) = \begin{pmatrix} a' & b' & \cdots & b' \\ b' & a' & \cdots & b' \\ \vdots & \vdots & \ddots & \vdots \\ b' & b' & \cdots & a' \end{pmatrix} \qquad (13)$$

The matrix elements $a'$ and $b'$ are complex, and the phase difference between them can be arbitrary. Accordingly, the matrix elements can be calculated as follows:

$$a' = e^{i\varsigma}\left(1 + \frac{1}{m}(-1 + e^{i\phi})\right) \qquad b' = e^{i\varsigma}\frac{1}{m}(-1 + e^{i\phi}) \qquad (14)$$

Expressing the exponents in trigonometric form, matrix elements can be written as:

$$a' = \frac{(n-1)\cos(\varsigma) + \cos(\phi + \varsigma) + i((n-1)\sin(\varsigma) + \sin(\phi + \varsigma))}{m} \qquad (15)$$

$$b' = \frac{-\cos(\varsigma) + \cos(\phi + \varsigma) - i(\sin(\varsigma) - \sin(\phi + \varsigma))}{m} \qquad (16)$$

The moduli and the phases of the matrix elements $a'$ and $b'$ are:

$$|a'| = \sqrt{2 + (-2 + m)m + 2(-1 + m)\cos(\phi)} \qquad (17)$$

$$\varphi_{a'} = -\operatorname{atan}\left(\frac{-\sec(\varsigma)\sin(\phi)}{(-1 + m)\cos(\varsigma) + \cos(\phi + \varsigma)} - \tan(\varsigma)\right) \qquad (18)$$

$$|b'| = \sqrt{\frac{2(1-\cos(\phi))}{m^2}} \tag{19}$$

$$\varphi_{b'} = \operatorname{atan}\left(-\cot\left(\frac{\phi}{2} + \varsigma\right)\right) \tag{20}$$

Equating the two matrices $C_0' = C_0$ (i.e. $a' = a$ and $b' = b$), the following expressions are obtained for traversing coin parameters by using Householder reflection

$$|a| = \sqrt{2 + (-2+m)m + 2(-1+m)\cos(\phi)} \tag{21}$$
$$= \frac{-2+m}{\sqrt{2 + (-2+m)m + 2(-1+m)\cos(2(\varphi_a - \varphi_b))}},$$

$$\varphi_a = -\operatorname{atan}\left(\frac{-\sec(\varsigma)\sin(\phi)}{(-1+r)\cos(\varsigma) + \cos(\phi+\varsigma)} - \tan(\varsigma)\right), \tag{22}$$

$$|b| = \sqrt{\frac{2(1-\cos(\phi))}{m^2}} = -\frac{2\cos(\varphi_a - \varphi_b)}{\sqrt{(-2+m)^2 + 4(-1+m)\cos(\varphi_a - \varphi_b)^2}}, \tag{23}$$

$$\varphi_b = \operatorname{atan}\left(-\cot\left(\frac{\phi}{2} + \varsigma\right)\right) \tag{24}$$

Therefore, the matrix elements of an arbitrary matrix with equal probability to go to each of the adjacent vertices can be constructed by both methods. If we search on structure where the size of coin $m = 2^n$, the necessary and sufficient parameters to find a good coin for traversing a regular graph (with equal probability for choosing any direction from all available, being at the current vertex) are the two angles $\varsigma, \phi$ and the number of qubits of the coin register $n$. The probability to find searched element depends on the coin. This can be written as follows:

$$p = p(\phi, \varsigma, n) \tag{25}$$

The hypercube is a particular case of such graph. All the above considerations also apply to it. For example, one coin that gives a very high probability to find the searched element is the Grover's coin. This coin can be obtained from equation (13) with the following values of the phases and the ket vector:

$$\varsigma = (2k_\varsigma + 1)\pi \qquad \phi = (2k_\phi + 1)\pi \qquad |\chi\rangle = \frac{1}{\sqrt{r}}\sum_{i=1}^{m}|i\rangle \tag{26}$$

Here $k_\varsigma$ and $k_\phi$ are arbitrary integers.

### 3.3. Benefits arising from the optimization of the coin operator for hypercube

The number of elements in register of the coin depends on the number of node register elements. The second parameter affects the probability of finding the solution by the following dependence:

$$p(\varsigma, \phi, n) = \frac{1}{2} - O\left(\frac{1}{2^m}\right) f(\varsigma, \phi) = \frac{1}{2} - O(2^{-2^n}) f(\varsigma, \phi) \tag{27}$$

This equation gives that the maximal probability that would be obtained at constant angles $\phi$ and $\varsigma$ depends on $2^{-2^n}$. The function $p(\varsigma, \phi, n)$ also depends on $\varsigma$ and $\phi$. At $\phi = \varsigma = \pi$ the Grover's coin is obtained. When $\phi = \varsigma = 0$ there is no random walk and the probability of finding the searched element is $2^{-2^n}$.

Although the addition of more qubits to the register of vertices and respectively - to the coin, will increase the probability of finding the required element, this work will focus on the two parameters in $f(\varsigma, \phi)$, the angles $\varsigma$ and $\phi$. Adding more qubits is a relatively complex task in the implementation of quantum computers. Reason behind this is that it is not only necessary to avoid the interactions between the qubits and the environment, but also to avoid the negative effects occurring within the quantum system itself. Properly selected phases, as will be seen in the next chapters, would always increase the stability of the algorithm against inaccurate coin parameters, regardless of the number of qubits. Increased stability would lead to a higher probability of obtaining the searched element in the experimental implementations.

Our goal is to find range of values of the phase parameters $\phi, \varsigma$ that give high probability to find the searched element $p = p(\phi, \varsigma, n)$, so the quantum algorithm is stable to changes in these parameters.

### 4. Numerical methods

Our approach to optimizing the QRWS algorithm is based on Monte Carlo and Machine Learning numerical methods. Here we will briefly discuss those methods and the way they are used.

#### 4.1. Monte Carlo

Monte Carlo (MC) methods are numerical stochastic methods [48], [49]. The numerical results are obtained by random sampling, and they could be used for generating random draws from a distribution. MC methods have application in optimization and numerical integration. Any problem that can be described by a mathematical function could be solved with this method by randomly generating the function's parameters and calculating an approximation of the function.

In this work a MC code have been written to simulate the quantum circuit of QRWS algorithm – with one-, two-, and three-qubit coin (see FIG. 1). For each of these coins with n qubits (n = 1, 2, and 3), our MC code calculates the probability to find the searched element $p(\phi, \varsigma, n)$ for a random pair of values of the angles $\phi, \varsigma = 0..2\pi$ and the number of steps for achieving maximum probability $k_{itr}(\phi, \varsigma, n)$. In the cases when the traversing coin is of one and two qubits, our code is relatively fast and we could generate enough datapoints to be able to use them in analyzing and optimizing the algorithm. But, as the dimension of the coin increases, the time for generating each datapoint grows exponentially. Due to this, we were able to create less data for n = 3 case, and we used Machine learning (ML) methods to fit the quantum model with a Deep neural network (DNN). The neural network (NN) has the potential to generate datapoints much faster than the actual quantum algorithm code and allows the study of more complex systems.

### 4.2. Machine Learning and Deep neural networks

Deep learning is a machine learning algorithm that consists of many layers of neurons, a feature that predisposes its use in fitting higher level features or fit more complex data. The DNNs input and output layers have a number of neurons equal to the input parameters and output elements correspondingly. The neural network is dense if all neurons between two sequential layers are connected. Each neuron has an activation function – linear, sigmoid, relu, selu, softmax, etc., that depends on the particular task for which the DNN is used. A representation of a deep neural network with L hidden layers and N neurons per layer is shown on FIG. 4.

The second step in our study was to build a proper DNN model from the collected MC datapoints. The ML methods are universal function approximators due to their ability to model the relationship between any given input and output data. The analysis further in the paper is based on this property. The MC datapoints for $n = 1, 2,$ and 3-qubit coins were used as training examples for fitting three Deep neural network models giving the probability $p(\phi, \varsigma, n)$ as a function of $\phi$ and $\varsigma$. The scheme of NN used is shown on FIG. 4. Finally, the models were applied in the analysis and optimization procedures as explained in the next sections.

### 4.3. Selection of proper neural network

Machine learning optimization is done by implementing Deep neural networks. The neural network models are created by using Keras deep learning API. We use sequential (feed forward) dense NNs. The optimal activation function for our problem is SELU (Scaled Exponential Linear Unit) for the input and the hidden layers, and sigmoid activation function for the output layer (see FIG. 4). The quality of the model is computed by squared error loss function and the training- and validation set loss functions are monitored as shown for a sample model on FIG. 5. We fit the DNN weight parameters with a fixed number of hidden layers and neurons per layer for a number of epochs. Each epoch the model is saved, and finally we select the model with a minimum value of the validation set loss function.

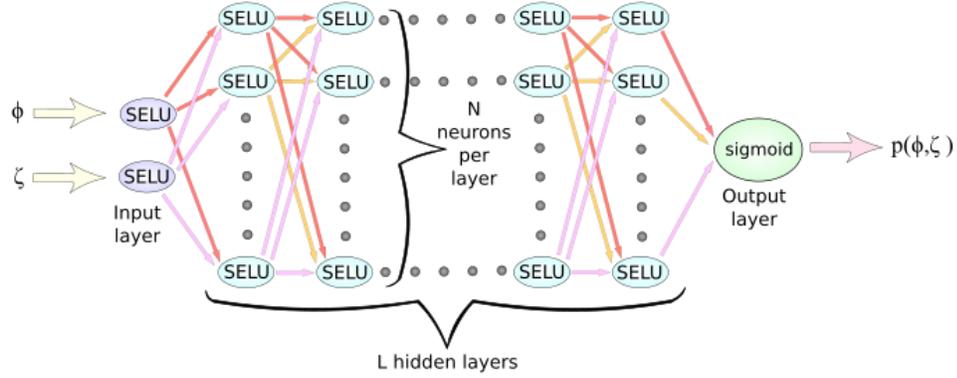

FIG. 4: *Scheme of the DNN used for fitting the probabilities $p(\phi, \varsigma)$ for a coin register consisting of n qubits $(p(\phi, \varsigma, n), n = 1, 2, 3)$.*

In order to find a suitable Deep learning model, we compare the quality for a number of different models fitting the one-, two-, and three-qubit coin MC data consisting of total 300 000 (15 000 in the 3-qubit coin case) datapoints. 80% of them are allocated for the training set and the rest - for the validation set. First, the minimum validation loss of each model is plotted for NNs with hidden layers up to 20 and neurons per layer from 5 to 30. Each model fitting code was run for a number of epochs using a batch size of 256 training examples and early stopping. In all three cases, we have plotted the validation loss as a function of the NN's number of layers and neurons per layer (the case of 3-qubit coin is shown on FIG. 6). The first few rows and columns with low number of hidden layers and neurons per layer have high validation loss (light colors) and are discarded. The same is the situation when we have too many layers. On the other hand, when more neurons are added, the data could be overfitted as the number of the model parameters becomes close to the number of the training examples. Even more, with more neurons the behavior of the model becomes more chaotic as we see more cells with very high loss (light color squares on the figure). In the next step, we find a few good deep learning models in areas with the lowest validation loss (dark colors). They are compared in detail by fitting procedure for a much higher number of epochs.

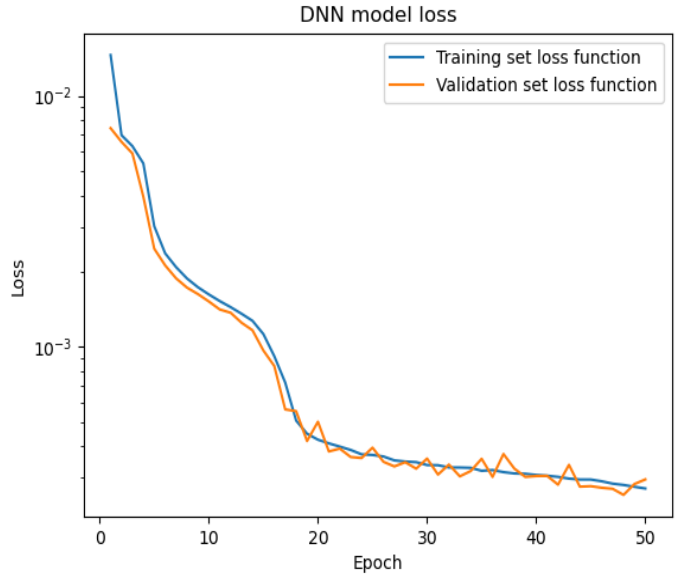

FIG. 5: *Training and validation sets loss functions of a test sequential DNN trained for fifty epochs.*

Finally, we have chosen the best model for each case (n = 1, 2, 3) with parameters: L = 7, 9, 7 hidden layers and N = 15, 13, 22 neurons per layer respectively.

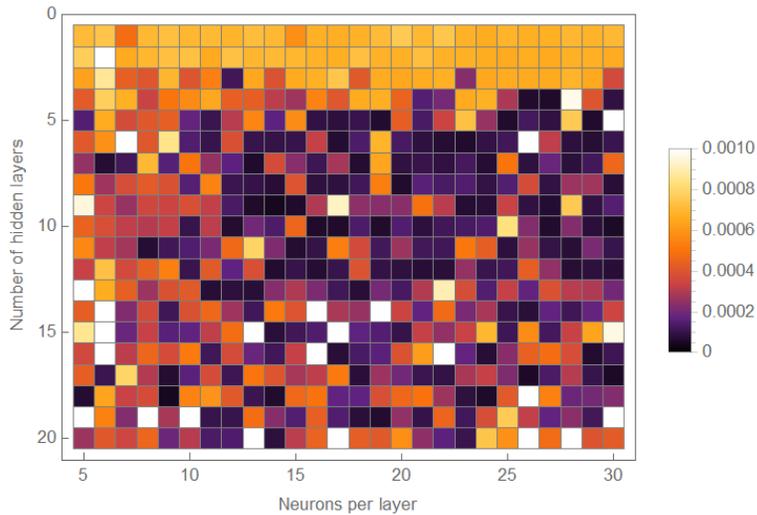

FIG. 6: *Validation loss as a function of number of layers L and neurons per layer N of the neural network. The darker squares correspond to a lower loss and better machine learning model.*

# 5. Numerical results

## 5.1. Monte Carlo simulations

To increase the stability of QRWS algorithm we could search for areas in the plane span by the coin angles $\phi$ and $\varsigma$ that give high probability to find solution when one or both of the parameters vary. In the best-case scenario a function can be found, as shown in Eq. (28) that gives relation between these phases. The number of qubits of the coin is prior known, so we will consider it as constant ($n = const$). In this case for fixed number of qubits we reduce the parameters of $p$ to one (29). Let $\phi_{max}$ be the value of the parameter $\phi$ at which maximum probability to find solution $p_{max}$ is obtained (30). In this scenario, if we define $\phi$ to vary in the interval $(\phi_{max} - \varepsilon_-, \phi_{max} + \varepsilon_+)$, then a specific function (28), giving the largest possible $\varepsilon_+ - \varepsilon_-$ that fulfill equation (31) is searched.

$$\varsigma = \varsigma(\phi) \tag{28}$$

$$p(\phi, \varsigma, n) \to p(\phi, \varsigma(\phi), n = const) \equiv p(\phi) \tag{29}$$

$$p(\phi_{max}) = p_{max} \tag{30}$$

$$p(\phi \in (\phi_{max} - \varepsilon_-, \phi_{max} + \varepsilon_+)) \cong p_{max} \tag{31}$$

The two traversing coin angles $\phi$ and $\varsigma$ will be used as coordinate axes in figures to be considered further in the chapter. Both angles are phase factors for the walk coin $C_0(\phi, \chi, \varsigma)$, and we could make numerical simulation of the coin with $\phi$ and $\varsigma$ in the interval $[0, 2\pi]$.

On FIG. 7 are shown the results of numerical simulations of quantum random walk search (QRWS) algorithm on hypercube using arbitrary values for both phases. The two coordinate axes correspond to the phases $\varsigma$ and $\phi$. Different colors correspond to different probabilities to find the required element after the number of iterations determined by Eq. (1), when using a coin based on a generalized Householder reflection and additional phase multiplier. The dark red corresponds to the highest probability of finding a solution and darker blue - to the lowest. In the case when $\phi = \varsigma = \pi$ the Grover's coin is obtained, and the probability to find a solution in case of coin consisting of two and three qubits are $p(\pi, \pi, 2) = 0.390625$ and $p(\pi, \pi, 3) = 0.434471$ respectively. When $n = 1$, the Grover's coin is not used because it gives a low probability $p(\pi, \pi, 1) = 0.25$.

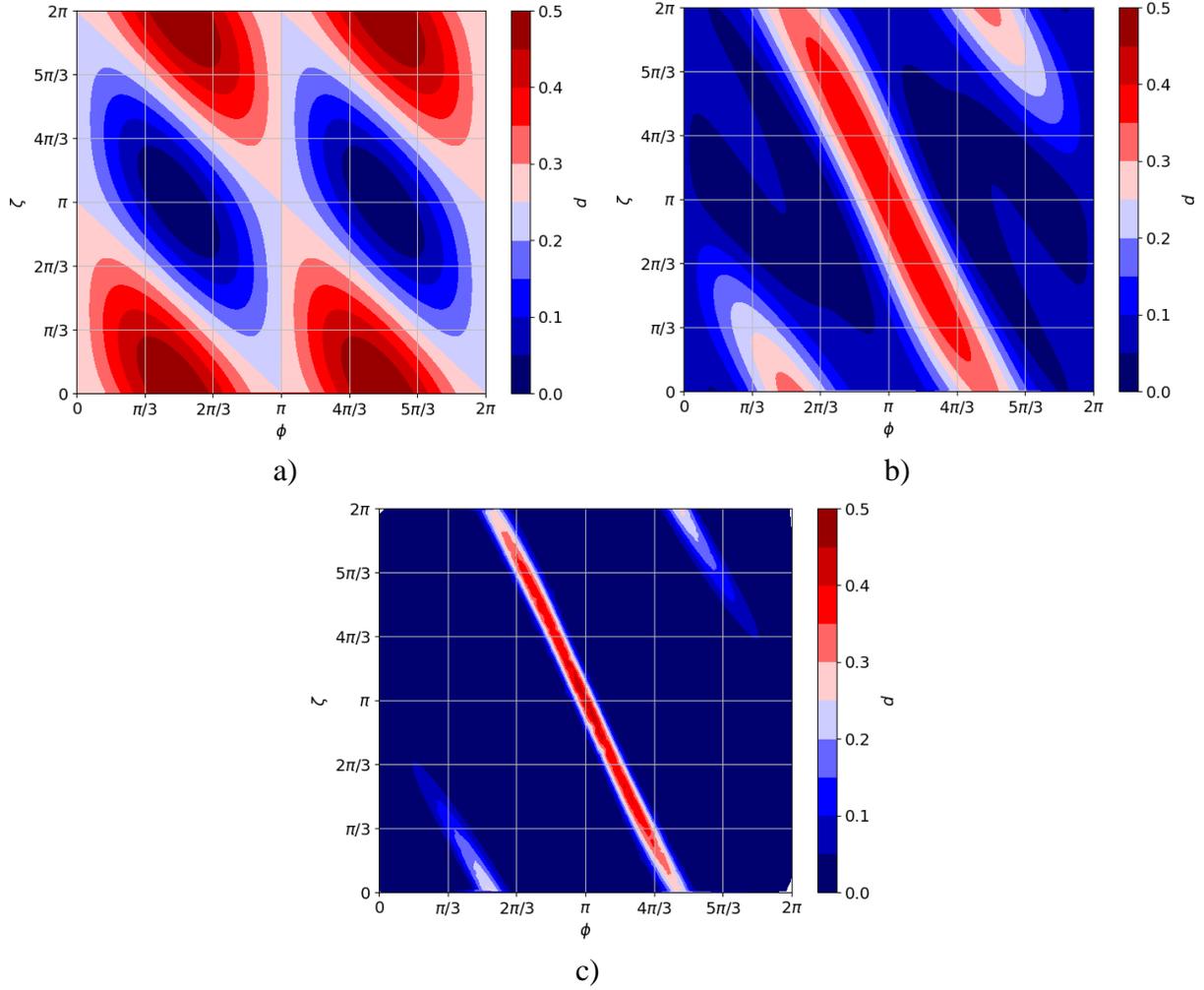

FIG. 7: *Results of simulations of the probability to find a solution with QRWS on hypercube in the case of one-, two- or three-qubit coin. The walk coin is built by generalized Householder reflection and additional phase multiplier. Two axes correspond to phases $\phi$ and $\varsigma$ used to construct the coin.*

It can be seen that for $n \geq 2$ the points with highest probability to find solution lie on parallel curves in the plane defined by $\phi$ and $\varsigma$. The curves have a period of $2\pi$ in angle $\phi$, due to the periodicity of the exponent.

These curves show there is relation between those phases, which we will look for as the functional dependence as assumed in equation (28). A linear function is the simplest possible approximation of this relation between the two phases. The formula for the linear function can be easily found using the coordinates of two consecutive maxima. We know that the coin obtained by the generalized Householder reflection, at $\varsigma = \pi$, achieves a very high probability when the other angle is $\phi = \pi$, which corresponds to a Grover's coin. The line passing through this point also passes through the point $\varsigma = 5\pi$ and $\phi = 3\pi$. This allows us to find the parallel lines which, in the approximation made, give the highest probability to find a solution when there are inaccuracies in the parameter $\phi$:

$$\varsigma = -2\phi + \pi + 2k_\phi \pi \qquad (32)$$

Therefore, if a coin is created by Householder reflection and inaccuracies in the phase $\phi$ are expected, then this linear function gives a higher probability $p(\phi, \varsigma, n)$, than all other possible lines.

The most unstable to parameter deviation line with high probability of finding a solution is a line perpendicular to it. An example of such line passing through the point $\varsigma = \pi$ and $\phi = \pi$ is:

$$\varsigma = 1/2\,\phi + \pi/2 \qquad (33)$$

Another line is when:

$$\varsigma = \pi \qquad (34)$$

A comparison of the probability of finding a stable solution to inaccuracies in the phase $\phi$ for the three lines is given on FIG. 8. The results are obtained from numerical simulations of the QRWS algorithm, where $\phi \epsilon [0, 2\pi]$. The two coordinate axes correspond to the phase $\phi$ and the probability of finding a solution $p$. Different colored curves correspond to different equations: red, magenta and teal curves correspond to simulation of equations (32), (33), and (34).

For the cases of a coin of one qubit, the maximum probability to find the searched element is achieved when the line is (32). The other two lines do not give a higher probability to measure the searched element than if the node register is measured without applying a QRWS. For the two- and three-qubit coin case it can be seen that the probability of finding a solution when there are inaccuracies in the phase $\phi$ is highest when using the line equation (32). In the case of a two-qubit coin, the line (32) also has a plateau with probability of finding the searched element equal to the probability obtained from the Grover's coin. For the case of a three-qubit coin, the maximum probability of finding a solution is at the point $\phi = \varsigma = \pi$, corresponding to the Grover's coin and is the same for the three lines.

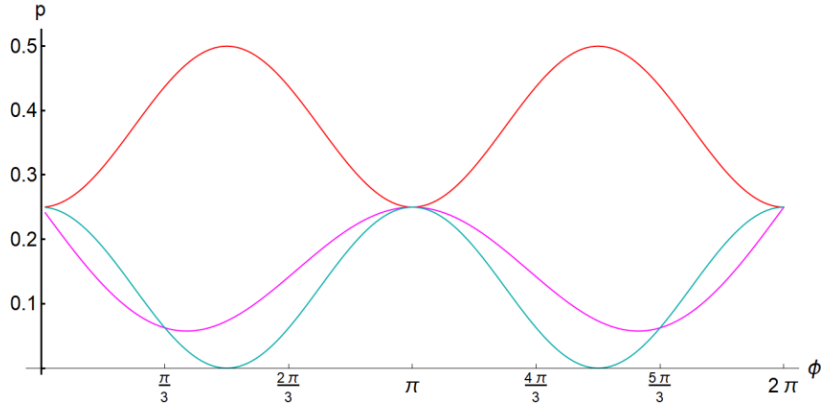

a)

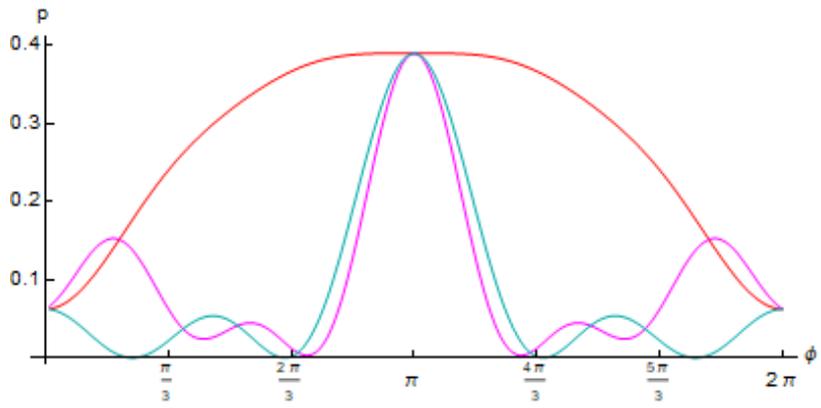

b)

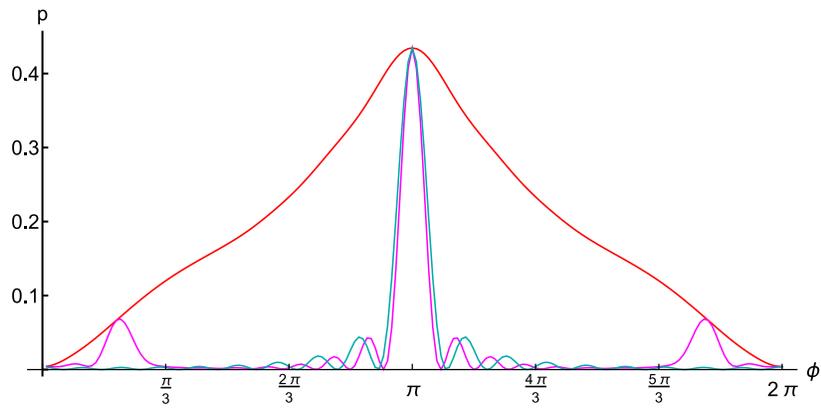

c)

FIG. 8: *The figure shows the probability of finding a solution by the QRWS algorithm with traversing coin build by generalized Householder reflection with phase $\phi \epsilon [0, 2\pi]$ and a global phase $\varsigma$ determined by equation of a line. The red curve corresponds to the line (32), magenta curve to (33), and teal curve to (34). Figure a) corresponds to one-qubit coin, figure b) to two-qubit coin $n = 2$, and figure c) shows results from the simulation with three-qubit coin $n = 3$.*

As the coin size increases, the linear approximation of the function becomes less appropriate. The reason for this is the decrease in the width of the high probability areas in the plane defined by $\varsigma$ and $\phi$, as can be seen in FIG. 7. The linear approximation (32) continues to perform better with increasing the number of qubits in the coin than the other lines discussed here. The highest probability of finding a solution lies on a curve described by a more complex function.

In the next chapter we will show the results obtained by machine learning for finding a better function $\varsigma = \varsigma(\phi)$, which will lead to a higher stability of the algorithm in finding the searched element.

### 5.2. Optimizing the probability to find solution by Machine Learning

The neural network model fitted to the Monte Carlo generated training datapoints is used to predict the values of $p(\phi, \varsigma, n)$ for any combination of the angles $\phi$ and $\varsigma$. Even more, it allows us to apply different optimization and fitting procedures to the DNN model, operations not directly applicable to the MC data. We have used standard minimization routines (differential evolution and simplicial homology global optimization with sobol sampling method) to search for the global extremum of the probability to find the element $p_{max}(\phi, \varsigma, n)$. Both methods are useful to deal with black box problems, where no prior knowledge of the functions is known. They do not require the optimization problem to be differentiable and thus do not impose any restrictions on the DNN model activation functions. On the other hand, the two routines are built on different mathematical bases, which allows obtaining independent results. It is important to note that this method for finding the global extremum of a neural network model is applicable to higher-dimension problems where no prior knowledge of the system is known.

From the optimization of the one-qubit coin QRWS algorithm (note that for 1-qubit QRWS algorithm there exist multiple global maxima for $p(\phi, \varsigma, n)$ when $\phi, \varsigma \epsilon [0, 2\pi]$), based on the constructed neural network model, we get $p_{max}$ very close to the expected theoretical maximum. See Table 1. Similar results were obtained for coin consisting of 3 qubits, where the maximum probability coincide with the one obtained by using the Grover's coin.

More interesting and unexpected results are obtained when the coin consist of two qubits – both minimization routines acting on the DNN model predicts coins that gives higher values of the probability in comparison to the probability given by the Grover's coin.

| | $\phi$ | $\varsigma$ | n | Method | Probability ($p_{max}$) |
|---|---|---|---|---|---|
| 1 | 4.590 | 0.319 | 1 | Sobol | 0.4932 |
| 2 | 4.704 | 6.283 | 1 | Differential evolution | 0.4996 |
| 3 | --- | --- | 1 | Theoretical maximum | 0.5000 |
| 4 | 3.145 | 3.143 | 2 | Sobol | 0.3907 |
| 5 | 2.764 | 3.986 | 2 | Differential evolution | 0.3915 |
| 6 | $\pi$ | $\pi$ | 2 | Grover's Coin | 0.3906 |
| 7 | 3.143 | 3.143 | 3 | Sobol | 0.4337 |
| 8 | 3.189 | 3.116 | 3 | Differential evolution | 0.4332 |
| 9 | $\pi$ | $\pi$ | 3 | Grover's Coin | 0.4345 |

*Table 1: Maximum probability $p_{max}$ found by using different methods and their respective values of $\phi$, $\varsigma$, and n.*

This discrepancy shows that further investigation should be conducted. We have studied in more detail the QRWS algorithm and found that, as shown in next section, that indeed for certain angles $(\phi, \varsigma)$, the maximum probability is slightly better than the expected one (when Grover's coin is used).

In the next paragraphs we analyze the QRWS algorithm in order to find the areas in $(\phi, \varsigma)$ plane, where it gives very high probability $p_{loc}$ close to the maximum probability to find the searched element $p_{loc}(\phi, \varsigma, n) \cong p_{max}$. We use both the Monte Carlo generated data and appropriate DNN models and compare the results. Our main goal is to improve the expression (32) $(\varsigma = -2\phi + \pi + 2k_\phi \pi, \phi \epsilon [0, 2\pi], k_\phi = 0, \pm 1, \pm 2, ...)$, with a function that gives high probability $p_{loc}(\phi, \varsigma(\phi), n)$ for larger interval of $\phi$, by including a small correction function and fitting it to the MC data and the neural network models. Better results were achieved by adding a sine function to line equation (32):

$$\varsigma = -2\phi + 3\pi + \alpha \sin(2\phi), \phi \epsilon [0, 2\pi]. \tag{35}$$

The optimal values of the coefficient α are given in Table 2 for a few coin registers with different number of qubits. They are found by fitting Eq. (35) to the Monte Carlo and the Neural network's data. In the cases with $n = 1, 2, 3$, both approaches lead to similar results. In the case of four-qubit coin only analysis of the DNN model is presented, as the numerical computation of QRWS algorithm by Monte Carlo simulations (for $n = 4$ or more) has extremely heavy hardware requirements.

| Walk coin | α (MC) | α (DNN) |
|---|---|---|
| 1-qubit | -0.467 | -0.535 |
| 2-qubit | -0.149 | -0.145 |
| 3-qubit | -0.202 | -0.204 |
| 4-qubit | --- | -0.248* |

Table 2: The values of the parameter α from Eq. (35) for different traversing coin with different number of qubits. The value in the last cell (marked with asterisk) is predicted from the DNN model as explained in Section 5.2.4.

We will review QRWS algorithm for 1-, 2-, 3-, and 4*-qubit (* explained in details in Section 5.2.4) coin separately and for each of them we stress on its individual peculiarities.

### 5. 2.1. One-qubit coin

It this case the walk coin has dimension two (one qubit), and it does not have any practical use, but it is important from a theoretical point of view to track the evolution of the algorithm when the coin size increases and to help build DNN model to predict the behavior of QRWS for bigger coin size. As seen on FIG. 7 a), minimums and maximums are chessly arranged, on first sight quite different from the pictures in the middle and on the right side of the figure. But it could be seen that by "squeezing" the picture along the line (32) the probability graph for the 1-qubit coin transforms to the 2-qubit one and further to the 3-qubit coin figure.

On FIG. 9 three lines giving the probability $p(\phi, \varsigma(\phi), 1)$ are shown. The red one is just the straight line from Eq. (32). The other two lines are obtained when the angle $\varsigma$ is given by Eq. (35), where the parameter α is $-1/(2\pi)$ (middle blue line) and $-2/\pi$ (green line). From the figure, it can be seen that both blue and green lines give higher probability to find the searched element for wider range of values of the angle $\phi$ than the line (32), derived in the previous section. $\alpha = -1/(2\pi)$ is a value that in our analysis proved to give better results for all coin sizes than the line $\varsigma = -2\phi + 3\pi, \phi \epsilon [0, 2\pi]$. The values of parameter $\alpha = -0.467$ and $\alpha = -0.535$, presented in Table 2, are extracted by fitting the MC data and the DNN model. Curves build by substituting these parameters in Eq. (11) overlap with the green line.

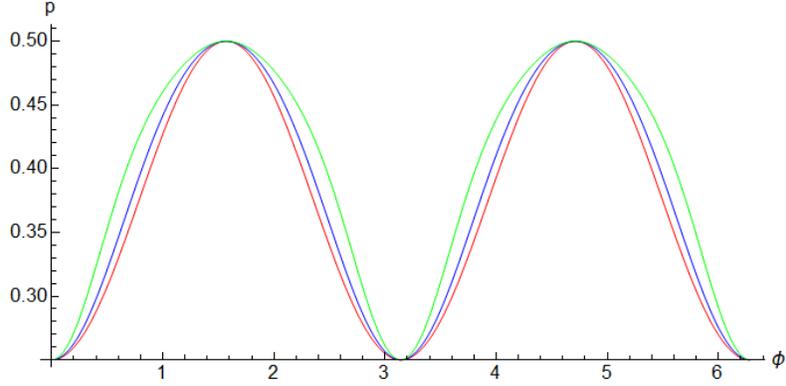

FIG. 9: *Probability to find the searched element by the QRWS algorithm for one-qubit coin register $p(\phi, \varsigma(\phi), 1)$. Each line represents different value of the parameter $\alpha$ in Eq. (35): $\alpha = -0.535$ (green), $\alpha = -1/(2\pi)$ (blue), $\alpha = 0$ (red).*

#### 5.2.2. Two-qubit coin

As mentioned above, by numerical analysis, we were able to identify points in the $\phi, \varsigma \in [0, 2\pi]$ plane with higher probability to find solution by the QRWS algorithm than the expected when a Grover's coin is used $p(\phi, \varsigma, 2) > p_{Grover}(\pi, \pi, 2) = 0.3906$. The results from a more detailed study are presented on FIG. 10 a). The red line shows the probability $p(\phi, \varsigma(\phi), 2)$ when $\varsigma = -2\phi + 3\pi$ and the green represents the probability when $\alpha = -1/(2\pi)$ in Eq. (35). In fact, the values $\alpha = -0.145$ (ML) and $\alpha = -0.149$ (MC) given in Table 2, obtained by Machine learning and Monte Carlo numerical analysis give slightly better results than the green line, but they are undistinguishable on the picture. On FIG. 10 b), on a larger scale, the area of the greatest interest is shown (see also FIG. 8 for comparison). For values $\phi \approx 4\pi/5$ and $\phi \approx 6\pi/5$ the probability $p(\phi, \varsigma, 2) \approx 0.392$ is slightly higher than the probability in the center of the graph corresponding to $p_{Grover}(\pi, \pi, 2)$.

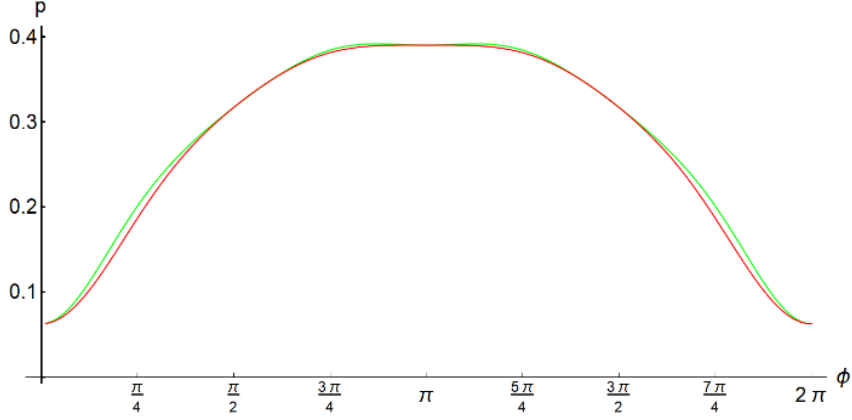

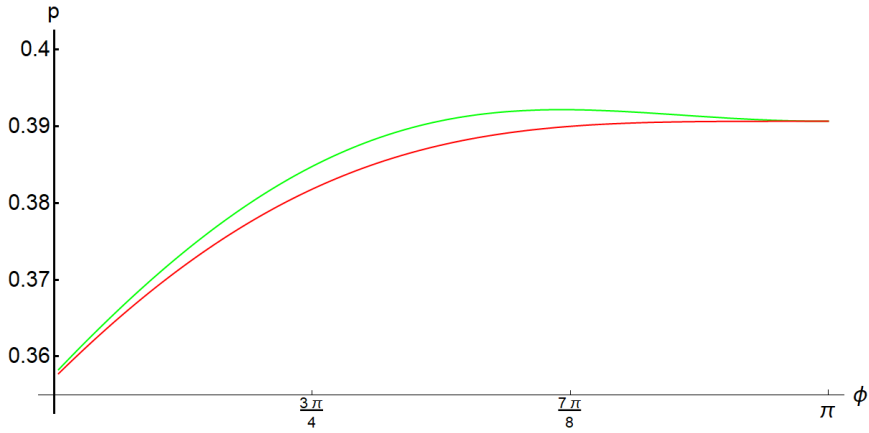

FIG. 10: *a) Probability to find the searched element by the QRWS algorithm for two-qubit walk coin $p(\phi, \varsigma(\phi), 2)$. Each line represents different value of the parameter $\alpha$ in Eq. (35) green – ($\alpha = -1/(2\pi)$ ) and $\alpha = 0$ (red). On b), the area of the greatest interest is shown on a larger scale.*

### 5.2.3. Three-qubit coin

In the two cases reviewed above (one and two-qubit coin), due to the large number of training examples we were able to generate, the results obtained independently by both Monte Carlo data and neural network models are identical. On the contrary, the number of datapoints generated for the three-qubit coin is much smaller (15 000) and in the optimization of the quantum algorithm we use the two methods together to check and complement the results we get from each of them in order to achieve the best optimization. To give an example of how one method could enhance the results from the other, we have plotted the probability $p(\phi, \varsigma, 3)$ by using the rough datapoints from the Monte Carlo code (FIG. 11 a)) and the probabilities predicted by our Machine learning model (FIG. 11 b)). It can be seen that the neural network fits very well the MC data and the figure is smoother. The Deep network model captures all the important characteristics of the function $p(\phi, \varsigma, 3)$ which justifies its use in the numerical analysis.

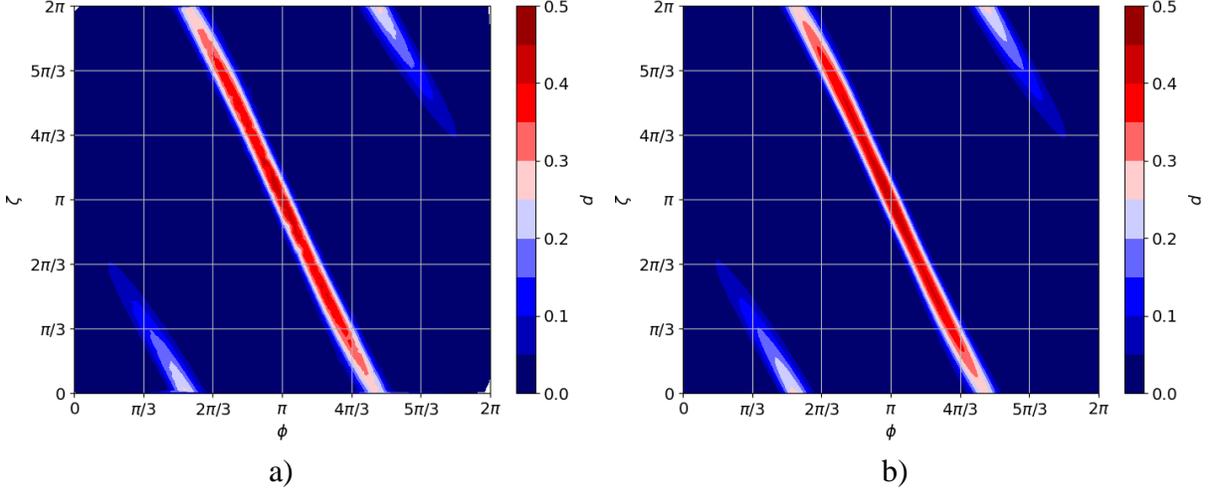

a) b)

FIG. 11: *Comparison between contour plots of the probability to find solution for the 3-qubit coin register $p(\phi, \varsigma, 3)$, $\phi, \varsigma \epsilon [0, 2\pi]$ from Monte Carlo data a) and Deep neural network model b).*

When the walk coin is of three qubits (n = 3), the area of high probability $p(\phi, \varsigma, 3)$ in the $\phi, \varsigma$ plane is close to the line $\varsigma = -2\phi + \pi + 2k_\phi \pi, \phi \epsilon [0, 2\pi], k_\phi = 0, \pm 1, \pm 2, ...$ It is narrower than in the case of two-qubit coin. We have fitted the points in this area to the expression (35) with both Monte Carlo data and even more successfully with the machine learning model. Nevertheless, at the end two methods gave similar results (see Table 2). On FIG. 12 are shown four lines of high probability $p(\phi, \varsigma, 3)$. Each of them is obtained by different value of the parameter α in the relation between the angles $\phi$ and $\varsigma$ given in (35). The red line gives the probability to find the solution without corrections to the line approximation in Eq. (32). The other three curves give better approximations with coefficients in front of the sine function $\alpha = -0.204$ (green), $\alpha = -1/(2\pi)$ (purple), and $\alpha = -1/(3\pi)$ (blue). The purple line is obtained with the same value of the parameter $\alpha = -1/(2\pi)$ as in the cases for smaller coin with n = 1, 2 qubits. It is clear that again it gives a very good approximation. The blue line is the first approximation obtained from the MC datapoints only. And the best one (green) was found when the Machine learning model was used. It gives about three times more stable QRWS algorithm to changes in $\phi$, than the line approximation (32). The value in the second column in Table 2 $\alpha = -0.202$ was obtained from the analysis of the MC data after some improvements were made dictated by the neural network model. In the case of three-qubit coin, the difference between the straight-line approximation and the approximations given in this chapter is more significant in comparison with the n = 2 case as seen from FIG. 10. We expect the optimizations done in this work to be even more important for higher walk coin size.

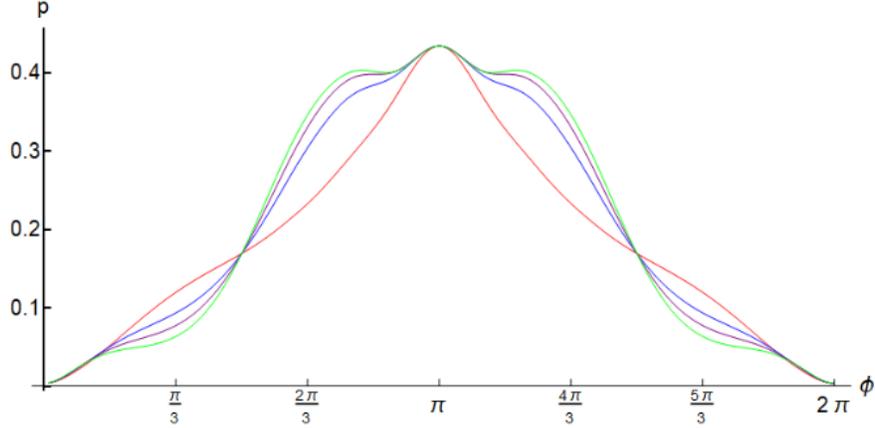

FIG. 12: *Probability to find the searched element by the QRWS algorithm for three-qubit coin register $p(\phi, \varsigma(\phi), 3)$. Each line represents different value of the parameter $\alpha$ in Eq. (35): $\alpha = -0.204$ (green), $\alpha = -1/(2\pi)$ (purple), $\alpha = -1/(3\pi)$ (blue), and in the case when $\alpha = 0$ (red) Eq. (35) coincide with Eq. (32).*

### 5.2.4 Optimizing the QRWS algorithm with bigger than three-qubit walk coin by Machine learning predictions

Direct implementation of the QRWS algorithm of 4-qubit walk coin, due to the large dimension of the gate operators, has very high demands to the computational resources and would require more powerful servers to run our Monte Carlo code. Even more, in order to optimize it in the way set out above, we need to generate a large amount of data. Thus, in the analysis of the four-qubit case we use a deep neural network to predict the behavior of the algorithm for different coin parameters. The DNN model is built by using the training examples generated by the MC code for 1, 2, and 3 qubit coin registers. It is known that the machine learning has excellent capabilities to interpolate data in the region it is trained on, but is not as efficient in extrapolating outside it. Thus, our predictions could only be confirmed when a comparison is made with an experimental data or more stringent numerical calculations.

On FIG. 13 the deep neural network used in the predictions of the probabilities $p(\phi, \varsigma, n), n = 4, 5, 6$ is shown. Unlike in the previous case, where we had two input parameters $\phi$ and $\varsigma$, here we add a third – the size of the walk coin with n qubits. Monte Carlo simulations' data of the QRWS algorithm for number of qubits of the coin register $n = 1, 2, 3$ is used to train the machine learning parameters. Few models were studied. All of them capture the main features of the quantum algorithm, but in the further study we have chosen the one with $L = 7$ hidden layers and $N = 24$ neurons per layer as it has shown best results in few categories – low validation loss, proper behavior even for higher coin register size, etc.

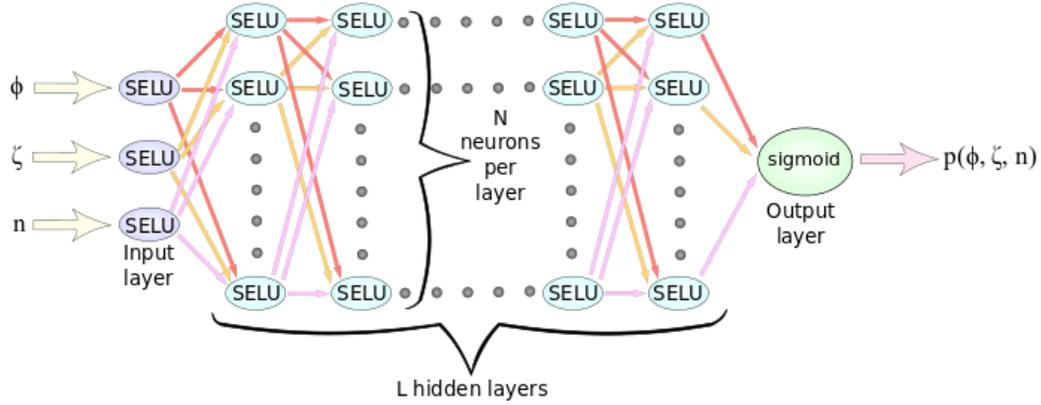

FIG. 13: *Scheme of the Deep neural network used for the predictions of the probabilities $p(\phi, \varsigma, n), n = 4, 5, 6$.*

The NN was used for prediction of the probabilities $p(\phi, \varsigma, n), n = 4, 5, 6$ for the walk coin phase parameters $\phi, \varsigma \epsilon [0, 2\pi]$ (see Eq. (13)). The results are shown on FIG. 14 for coin register consisting of 4, 5, and 6 qubits. A brief comparison with the corresponding plots for lower coin sizes from FIG. 7, shows that the behavior seen there can be observed here as well. Namely, the thinning of the central red stripe corresponding to high values for the probability $p(\phi, \varsigma, n)$ and the reduction of the lateral stripes with a high probability. On the other hand, the symmetry for lower- and higher than $\pi$ values of $\phi$ is broken. This is even more clearly visible when the machine learning model is used to predict the probability for bigger coin with register consisting of more qubits $n = 5, 6$.

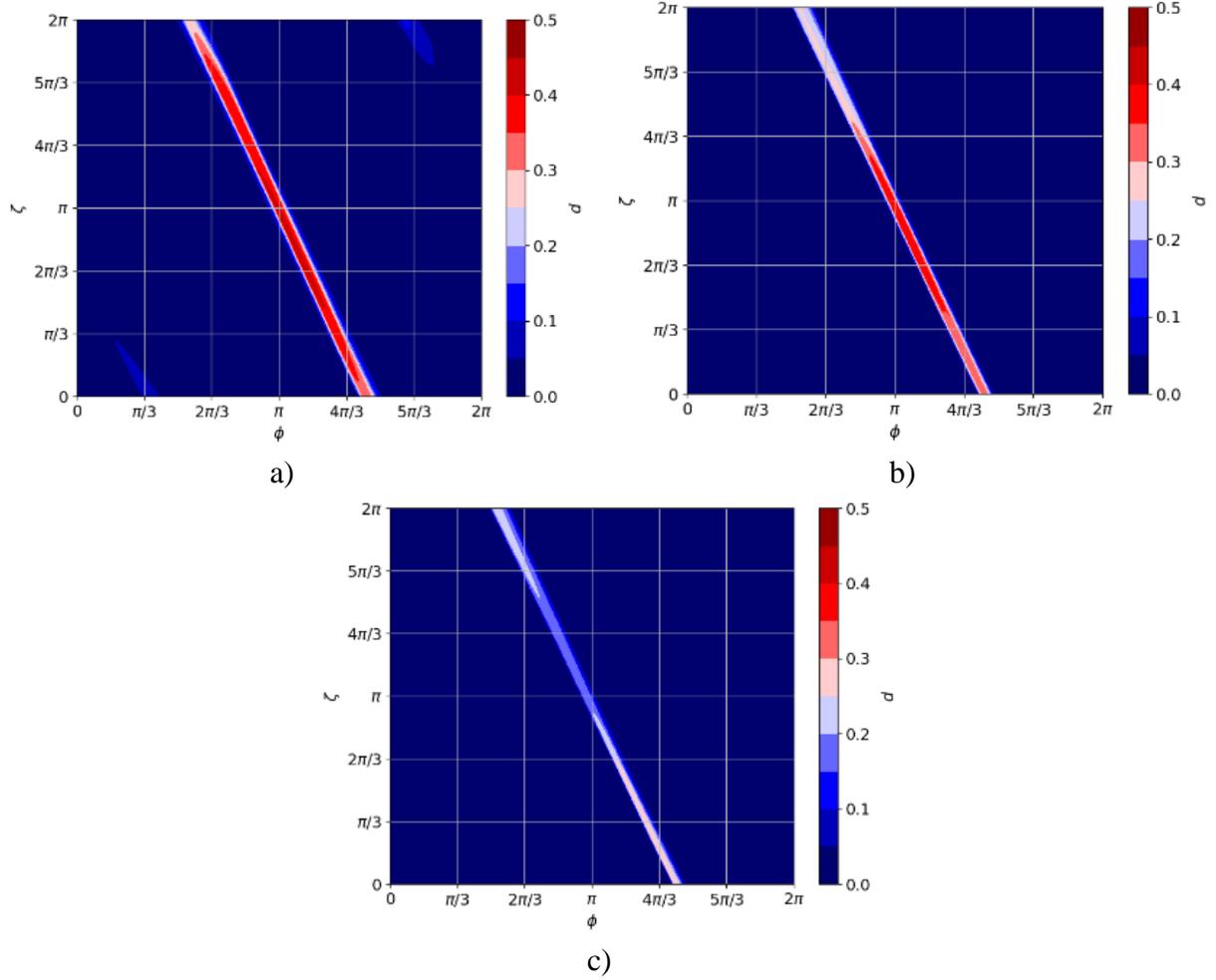

FIG. 14: *Predicted probabilities $p(\phi, \varsigma, n), n = 4, 5, 6$ (figures a), b), and c) accordingly) from the DNN model plotted for all values of the phase angles $\phi, \varsigma \epsilon [0, 2\pi]$.*

      In the case of four-qubit walk coin only, we have conducted a more detailed study. Fitting procedures were applied to the DNN model, and the obtained result for the fitting parameter $\alpha$ from Eq. (35), giving the prognosis line with maximum probability $p(\phi, \varsigma(\phi), 4)$ in the $(\phi, \varsigma)$ plane, is $\alpha = -0.248$. On FIG. 15 are shown four lines representing the probability to find the searched element by the QRWS algorithm for four-qubit coin register $p(\phi, \varsigma(\phi), 4)$ for three chosen values of the parameter $\alpha$: $\alpha = -0.248$ (green), $\alpha = -1/(2\pi)$, (purple), $\alpha = 0$ (red), and when $\varsigma = \pi$ (teal). The narrower teal line could be obtained from the leftmost picture on FIG. 14 when $\varsigma = \pi$. It is clear that a Quantum random walk search algorithm with a walk coin constructed in this way will be a way more unstable to experimental inaccuracies of the phase $\phi$. More stable will be a realization of the algorithm when the simple linear relation between the phases $\phi$ and $\varsigma$ (Eq. (32)) is realized (red line). An even better solution is given by the purple curve that is constructed when in Eq. (35), the parameter $\alpha$ is $-1/(2\pi)$. This solution is quite general as it proves to excel in comparison to the simple line approximation for all studied walk coin sizes. The best fit was obtained by Machine learning methods when $\alpha = -0.248$ – the green line. It is wider in comparison to all other fits to the expression (35) and thus corresponds to the most stable implementation of the QRWS algorithm on a hypercube.

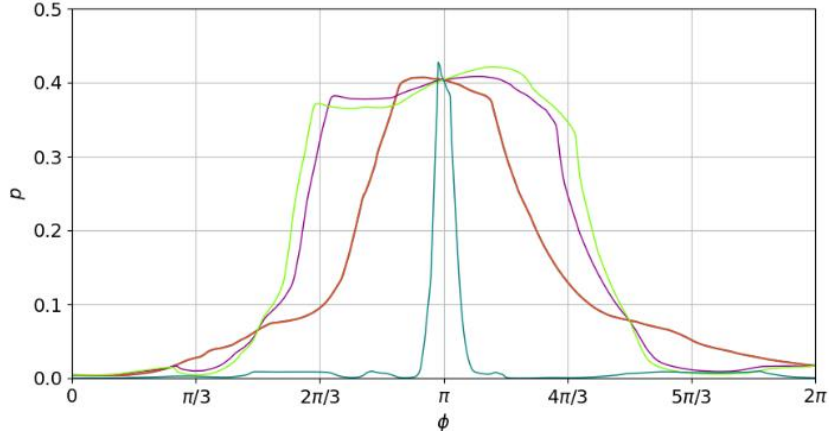

FIG. 15: *Probability to find the searched element by the QRWS algorithm for four-qubit coin register $p(\phi, \varsigma(\phi), 4)$. Each line represents different value of the parameter $\alpha$ in Eq. (35): $\alpha = -0.248$ (green), $\alpha = -1/(2\pi)$ (purple), $\alpha = 0$ (red) and when $\varsigma = \pi$ (teal). The curves are extracted from our neural network model and represent the high probability points from the red straps on the leftmost picture on FIG. 14.*

As explained above, although the DNN model predicts some unphysical asymmetries, it can correctly simulate the behavior of the quantum algorithm and valuable information from applying numerical methods to the model can be extracted.

Our analysis in this chapter shows that, if linear dependance Eq. (32) between phases is achieved, this will increase the stability of algorithm when phase $\phi$ can't be made accurate enough, even with bigger coin register size. Better performance could be obtained if Eq. (35) is used instead. However, the last realization is more demanding to implement.

### 6. Remarks on experimental implementation of walk coin described above

Building an effective walk coin by Householder reflection and phase gate, as described in this work, requires:

1) To be able to build generalized Householder reflection effectively, with an arbitrary phase
2) To obtain phase gate with a given phase
3) To be able to make desired correlation between these phases, shown in this work.

Householder reflection can be implemented effectively on some of the physical systems suitable for quantum computer realizations. Example of such system is the ion trap. More detailed information of this can be seen in [17], [18] and [19]. Here we will give only brief remarks:

Ions in the trap have to be cooled, and their motion on the axis perpendicular to the axis connecting them should be neglectable. If coupling fields coincide in time, then Morris-Shore transformation [64] can be used to change the basis. In this basis the reflection parameters can be controlled by using parameters of the laser pulse. $|\chi\rangle$ is determined by the Rabi frequencies of the system - lasers and the corresponding ions they control. Phase $\phi$ of the generalized Householder reflection is determined by the pulse shape and detuning. An equation for the phase in case of hyperbolic-secant pulse is shown in [19].

Phase gate can be constructed by use of STIRAP [65] like setting for tripod system in the way as is shown in [66]. Phases generated in this way depend on the shapes of Stokes and pump pulses, the phase differences between them, and the relative field strengths.

Functional dependance between the phases as is in equation (32) or (35) could be done by using beam splitter to divide laser light into two beams. The first beam should be used to implement the generalized Householder reflection as is described above. The second beam should be used to construct the phase gate. We suppose that those methods can be used to build the desired relations between phases, but more investigation is needed.

## 7. Conclusion

In this work is investigated the probability to find solution of quantum random walk search algorithm depending on a walk coin constructed by one Generalized Householder reflection and one phase multiplier for all states. Both Generalized Householder reflection and phase gate can be constructed in experimental implementations of quantum computer like the ion traps. The coin parameters used in the optimization are derived analytically – the number of qubits in the coin, the Generalized Householder reflection phase and the phase multiplier for all states. Monte Carlo and Machine learning methods are applied to find practically useful parametric relation between these phases. It is shown that if an appropriate dependance between both phases is achieved, this will greatly increase the stability of the Quantum random walk search algorithm to inaccuracies in the phases. The cases of 1,2, and 3-qubit walk coin are studied in detail as an example. For each of these cases a simple linear function is found that could have easier implantation. Also, more complex function is derived but it gives higher stability of the quantum algorithm. In the case of two-qubit coin, phases that give slightly higher probability to find solution than Grover's coin were found. A deep neural network was used to make predictions for 4-qubit coin. Relations between the phases are found in this case too, but the results should be confirmed by conducting more in-depth numerical calculations. The numerical calculations for 1, 2, and 3 qubits of the coin, and the neural network analysis for 4 and more qubits show that the results for building the walk coin presented in this work have to be valid even for larger coin register size.

## 8. Acknowledgments

The work on this paper was supported by the Bulgarian National Science Fund under Grant KP-06-M48/2 from 26.11.2020.

References:

[1] M. A. Nielsen and I. L. Chuang, *Quantum computation and quantum information*. Cambridge: Cambridge Univ. Press, 2007.

[2] G. van Assche, *Quantum Cryptography and Secret-Key Distillation*. Cambridge University Press, 2006.

[3] C. Gross and I. Bloch, "Quantum simulations with ultracold atoms in optical lattices," *Science*, vol. 357, no. 6355, pp. 995–1001, Sep. 2017, doi: 10.1126/science.aal3837.

[4] E. Bernstein and U. Vazirani, "Quantum Complexity Theory," *SIAM J. Comput.*, vol. 26, no. 5, pp. 1411–1473, Oct. 1997, doi: 10.1137/S0097539796300921.

[5] A. W. Burks and A. R. Burks, "First General-Purpose Electronic Computer," *IEEE Annals of the History of Computing*, vol. 3, no. 04, pp. 310–389, Oct. 1981, doi: 10.1109/MAHC.1981.10043.

[6] F. Arute *et al.*, "Quantum supremacy using a programmable superconducting processor," *Nature*, vol. 574, no. 7779, pp. 505–510, Oct. 2019, doi: 10.1038/s41586-019-1666-5.

[7] J. Porter, "Google confirms 'quantum supremacy' breakthrough," *The Verge*, Oct. 23, 2019. https://www.theverge.com/2019/10/23/20928294/google-quantum-supremacy-sycamore-computer-qubit-milestone (accessed Apr. 04, 2021).

[8] J. A. Jones and M. Mosca, "Implementation of a quantum algorithm on a nuclear magnetic resonance quantum computer," *J. Chem. Phys.*, vol. 109, no. 5, pp. 1648–1653, Jul. 1998, doi: 10.1063/1.476739.

[9] G.-Q. Liu and X.-Y. Pan, "Quantum information processing with nitrogen–vacancy centers in diamond," *Chinese Phys. B*, vol. 27, no. 2, p. 020304, Feb. 2018, doi: 10.1088/1674-1056/27/2/020304.

[10] D. Loss and D. P. DiVincenzo, "Quantum computation with quantum dots," *Phys. Rev. A*, vol. 57, no. 1, pp. 120–126, Jan. 1998, doi: 10.1103/PhysRevA.57.120.

[11] D. Kielpinski, C. Monroe, and D. J. Wineland, "Architecture for a large-scale ion-trap quantum computer," *Nature*, vol. 417, no. 6890, pp. 709–711, Jun. 2002, doi: 10.1038/nature00784.

[12] A. Derevianko and C. C. Cannon, "Quantum computing with magnetically interacting atoms," *Phys. Rev. A*, vol. 70, no. 6, p. 062319, Dec. 2004, doi: 10.1103/PhysRevA.70.062319.

[13] V. K. Kaplunenko and A. V. Ustinov, "Experimental test of a superconducting digital interface for vortex qubits," *Eur. Phys. J. B*, vol. 38, no. 1, pp. 3–8, Mar. 2004, doi: 10.1140/epjb/e2004-00091-3.

[14] F. Jazaeri, A. Beckers, A. Tajalli, and J. Sallese, "A Review on Quantum Computing: From Qubits to Front-end Electronics and Cryogenic MOSFET Physics," pp. 15–25, doi: 10.23919/MIXDES.2019.8787164.

[15] D. P. DiVincenzo, "The Physical Implementation of Quantum Computation," *Fortschritte der Physik*, vol. 48, no. 9-11, pp. 771–783, Sep. 2000, doi: 10.1002/1521-3978(200009)48:9/11<771::AID-PROP771>3.0.CO;2-E.

[16] S. Resch and U. R. Karpuzcu, "Quantum Computing: An Overview Across the System Stack," *arXiv:1905.07240 [quant-ph]*, Oct. 2019, Accessed: Apr. 04, 2021. [Online]. Available: http://arxiv.org/abs/1905.07240.

[17] P. A. Ivanov, E. S. Kyoseva, and N. V. Vitanov, "Engineering of arbitrary $\mathrm{U}(N)$ transformations by quantum Householder reflections," *Phys. Rev. A*, vol. 74, no. 2, p. 022323, Aug. 2006, doi: 10.1103/PhysRevA.74.022323.

[18] P. A. Ivanov and N. V. Vitanov, "Synthesis of arbitrary unitary transformations of collective states of trapped ions by quantum Householder reflections," *Phys. Rev. A*, vol. 77, no. 1, p. 012335, Jan. 2008, doi: 10.1103/PhysRevA.77.012335.

[19] S. S. Ivanov, H. S. Tonchev, and N. V. Vitanov, "Time-efficient implementation of quantum search with qudits," *Phys. Rev. A*, vol. 85, no. 6, p. 062321, Jun. 2012, doi: 10.1103/PhysRevA.85.062321.

[20] "Quantum Takes Flight: Moving from Laboratory Demonstrations to Building Systems," *IBM Research Blog*, Jan. 08, 2020. https://www.ibm.com/blogs/research/2020/01/quantum-volume-32/ (accessed Apr. 02, 2021).

[21] "Intel Labs is producing quantum processors in Oregon and doing system-level engineering that targets production-level quantum computing within ten years," *Intel*. https://www.intel.com/content/www/us/en/research/quantum-computing.html (accessed Apr. 02, 2021).

[22] J. A. Buchmann, D. Butin, F. Göpfert, and A. Petzoldt, "Post-Quantum Cryptography: State of the Art," in *The New Codebreakers: Essays Dedicated to David Kahn on the Occasion of His 85th Birthday*, P. Y. A. Ryan, D. Naccache, and J.-J. Quisquater, Eds. Berlin, Heidelberg: Springer, 2016, pp. 88–108.

[23] D. J. Bernstein and T. Lange, "Post-quantum cryptography," *Nature*, vol. 549, no. 7671, Art. no. 7671, Sep. 2017, doi: 10.1038/nature23461.

[24] N. Shenvi, J. Kempe, and K. B. Whaley, "Quantum random-walk search algorithm," *Phys. Rev. A*, vol. 67, no. 5, p. 052307, May 2003, doi: 10.1103/PhysRevA.67.052307.

[25] A. M. Childs and J. Goldstone, "Spatial search by quantum walk," *Phys. Rev. A*, vol. 70, no. 2, p. 022314, Aug. 2004, doi: 10.1103/PhysRevA.70.022314.

[26] A. M. Childs, S. Kimmel, and R. Kothari, "The Quantum Query Complexity of Read-Many Formulas," in *Algorithms – ESA 2012*, Berlin, Heidelberg, 2012, pp. 337–348, doi: 10.1007/978-3-642-33090-2_30.

[27] M. Ben-Or and A. Hassidim, "Quantum Search in an Ordered List via Adaptive Learning," *arXiv:quant-ph/0703231*, Nov. 2007, Accessed: Apr. 04, 2021. [Online]. Available: http://arxiv.org/abs/quant-ph/0703231.

[28]   R. Franco, "Quantum Amplitude Amplification Algorithm: An Explanation of Availability Bias," in *Quantum Interaction*, Berlin, Heidelberg, 2009, pp. 84–96, doi: 10.1007/978-3-642-00834-4_9.

[29]   G. Brassard, P. Hoyer, M. Mosca, and A. Tapp, "Quantum Amplitude Amplification and Estimation," *arXiv:quant-ph/0005055*, vol. 305, pp. 53–74, 2002, doi: 10.1090/conm/305/05215.

[30]   H. S. Tonchev and N. V. Vitanov, "Quantum phase estimation and quantum counting with qudits," *Phys. Rev. A*, vol. 94, no. 4, p. 042307, Oct. 2016, doi: 10.1103/PhysRevA.94.042307.

[31]   L. Ruiz-Perez and J. C. Garcia-Escartin, "Quantum arithmetic with the quantum Fourier transform," *Quantum Inf Process*, vol. 16, no. 6, p. 152, Apr. 2017, doi: 10.1007/s11128-017-1603-1.

[32]   V. M. Kendon, "A random walk approach to quantum algorithms," *Philosophical Transactions of the Royal Society A: Mathematical, Physical and Engineering Sciences*, vol. 364, no. 1849, pp. 3407–3422, Dec. 2006, doi: 10.1098/rsta.2006.1901.

[33]   E. Farhi and S. Gutmann, "Quantum computation and decision trees," *Phys. Rev. A*, vol. 58, no. 2, pp. 915–928, Aug. 1998, doi: 10.1103/PhysRevA.58.915.

[34]   Y. Aharonov, L. Davidovich, and N. Zagury, "Quantum random walks," *Phys. Rev. A*, vol. 48, no. 2, pp. 1687–1690, Aug. 1993, doi: 10.1103/PhysRevA.48.1687.

[35]   B. C. Travaglione and G. J. Milburn, "Implementing the quantum random walk," *Phys. Rev. A*, vol. 65, no. 3, p. 032310, Feb. 2002, doi: 10.1103/PhysRevA.65.032310.

[36]   A. Ambainis, J. Kempe, and A. Rivosh, "Coins Make Quantum Walks Faster," *arXiv:quant-ph/0402107*, Feb. 2004, Accessed: Apr. 04, 2021. [Online]. Available: http://arxiv.org/abs/quant-ph/0402107.

[37]   A. Di Crescenzo, C. Macci, B. Martinucci, and S. Spina, "Analysis of random walks on a hexagonal lattice," *IMA J Appl Math*, vol. 84, no. 6, pp. 1061–1081, Dec. 2019, doi: 10.1093/imamat/hxz026.

[38]   J. Kempe, "Quantum Random Walks Hit Exponentially Faster," *arXiv:quant-ph/0205083*, May 2002, Accessed: Apr. 04, 2021. [Online]. Available: http://arxiv.org/abs/quant-ph/0205083.

[39]   A. Ambainis, "Quantum Walk Algorithm for Element Distinctness," *SIAM J. Comput.*, vol. 37, no. 1, pp. 210–239, Jan. 2007, doi: 10.1137/S0097539705447311.

[40]   F. Magniez, M. Santha, and M. Szegedy, "Quantum Algorithms for the Triangle Problem," *SIAM J. Comput.*, vol. 37, no. 2, pp. 413–424, Jan. 2007, doi: 10.1137/050643684.

[41]   A. Ambainis, A. M. Childs, B. W. Reichardt, R. Špalek, and S. Zhang, "Any AND-OR Formula of Size N Can Be Evaluated in Time $N^{1/2+o(1)}$ on a Quantum Computer," *SIAM J. Comput.*, vol. 39, no. 6, pp. 2513–2530, Jan. 2010, doi: 10.1137/080712167.

[42] D. Koch and M. Hillery, "Finding paths in tree graphs with a quantum walk," *Phys. Rev. A*, vol. 97, no. 1, p. 012308, Jan. 2018, doi: 10.1103/PhysRevA.97.012308.

[43] V. Potoček, A. Gábris, T. Kiss, and I. Jex, "Optimized quantum random-walk search algorithms on the hypercube," *Phys. Rev. A*, vol. 79, no. 1, p. 012325, Jan. 2009, doi: 10.1103/PhysRevA.79.012325.

[44] B. Hein and G. Tanner, "Quantum search algorithms on a regular lattice," *Phys. Rev. A*, vol. 82, no. 1, p. 012326, Jul. 2010, doi: 10.1103/PhysRevA.82.012326.

[45] A. Patel and K. S. Raghunathan, "Search on a fractal lattice using a quantum random walk," *Phys. Rev. A*, vol. 86, no. 1, p. 012332, Jul. 2012, doi: 10.1103/PhysRevA.86.012332.

[46] H. Tonchev, "Alternative Coins for Quantum Random Walk Search Optimized for a Hypercube," *Journal of Quantum Information Science*, vol. 05, no. 01, p. 6, Mar. 2015, doi: 10.4236/jqis.2015.51002.

[47] Y.-C. Zhang, W.-S. Bao, X. Wang, and X.-Q. Fu, "Optimized quantum random-walk search algorithm for multi-solution search," *Chinese Phys. B*, vol. 24, no. 11, p. 110309, Nov. 2015, doi: 10.1088/1674-1056/24/11/110309.

[48] J. Hammersley, *Monte Carlo Methods*. Springer Science & Business Media, 2013.

[49] M. H. Kalos and P. A. Whitlock, *Monte Carlo Methods*. John Wiley & Sons, 2009.

[50] W. Janke, "Monte Carlo Simulations of Spin Systems," in *Computational Physics: Selected Methods Simple Exercises Serious Applications*, K. H. Hoffmann and M. Schreiber, Eds. Berlin, Heidelberg: Springer, 1996, pp. 10–43.

[51] V. Blobel, "An Unfolding Method for High Energy Physics Experiments," *arXiv:hep-ex/0208022*, Aug. 2002, Accessed: Apr. 04, 2021. [Online]. Available: http://arxiv.org/abs/hep-ex/0208022.

[52] D. J. Earl and M. W. Deem, "Monte Carlo Simulations," in *Molecular Modeling of Proteins*, A. Kukol, Ed. Totowa, NJ: Humana Press, 2008, pp. 25–36.

[53] B. F. J. Manly, *Randomization, Bootstrap and Monte Carlo Methods in Biology, Third Edition*. CRC Press, 2006.

[54] T. Mitchell, *Machine Learning*. McGraw-Hill Education, 1997.

[55] A. Coates, A. Ng, and H. Lee, "An Analysis of Single-Layer Networks in Unsupervised Feature Learning," in *Proceedings of the Fourteenth International Conference on Artificial Intelligence and Statistics*, Jun. 2011, pp. 215–223, Accessed: Apr. 24, 2021. [Online]. Available: http://proceedings.mlr.press/v15/coates11a.html.

[56] A. Singh, N. Thakur, and A. Sharma, "A review of supervised machine learning algorithms," in *2016 3rd International Conference on Computing for Sustainable Global Development (INDIACom)*, Mar. 2016, pp. 1310–1315.

[57] X. Xu, L. Zuo, and Z. Huang, "Reinforcement learning algorithms with function approximation: Recent advances and applications," *Information Sciences*, vol. 261, pp. 1–31, Mar. 2014, doi: 10.1016/j.ins.2013.08.037.

[58] X.-C. Yang, M.-H. Yung, and X. Wang, "Neural-network-designed pulse sequences for robust control of singlet-triplet qubits," *Phys. Rev. A*, vol. 97, no. 4, p. 042324, Apr. 2018, doi: 10.1103/PhysRevA.97.042324.

[59] Z. Zhu, B. Dong, H. Guo, T. Yang, and Z. Zhang, "Fundamental band gap and alignment of two-dimensional semiconductors explored by machine learning," *Chinese Phys. B*, vol. 29, no. 4, p. 046101, Mar. 2020, doi: 10.1088/1674-1056/ab75d5.

[60] J. M. Arrazola, T. Kalajdzievski, C. Weedbrook, and S. Lloyd, "Quantum algorithm for nonhomogeneous linear partial differential equations," *Phys. Rev. A*, vol. 100, no. 3, p. 032306, Sep. 2019, doi: 10.1103/PhysRevA.100.032306.

[61] J. Bang, J. Ryu, S. Yoo, M. Pawłowski, and J. Lee, "A strategy for quantum algorithm design assisted by machine learning," *New J. Phys.*, vol. 16, no. 7, p. 073017, Jul. 2014, doi: 10.1088/1367-2630/16/7/073017.

[62] L. Cincio, Y. Subaşı, A. T. Sornborger, and P. J. Coles, "Learning the quantum algorithm for state overlap," *New J. Phys.*, vol. 20, no. 11, p. 113022, Nov. 2018, doi: 10.1088/1367-2630/aae94a.

[63] S. Hoyer, "Quantum random walk search on satisfiability problems," 2008, Accessed: Apr. 04, 2021. [Online]. Available: https://scholarship.tricolib.brynmawr.edu/handle/10066/10450.

[64] B. W. Shore, "Two-state behavior in N-state quantum systems: The Morris–Shore transformation reviewed," *Journal of Modern Optics*, vol. 61, no. 10, pp. 787–815, Jun. 2014, doi: 10.1080/09500340.2013.837205.

[65] N. V. Vitanov, A. A. Rangelov, B. W. Shore, and K. Bergmann, "Stimulated Raman adiabatic passage in physics, chemistry, and beyond," *Rev. Mod. Phys.*, vol. 89, no. 1, p. 015006, Mar. 2017, doi: 10.1103/RevModPhys.89.015006.

[66] D. Møller, L. B. Madsen, and K. Mølmer, "Geometric phase gates based on stimulated Raman adiabatic passage in tripod systems," *Phys. Rev. A*, vol. 75, no. 6, p. 062302, Jun. 2007, doi: 10.1103/PhysRevA.75.062302.